\documentclass{article}
\usepackage{arxiv}
\usepackage[utf8]{inputenc} 
\usepackage[T1]{fontenc}    
\usepackage{hyperref}       
\usepackage{url}            
\usepackage{booktabs}       
\usepackage{amsfonts}       
\usepackage{nicefrac}       
\usepackage{microtype}      
\usepackage{lipsum}
\usepackage{graphicx}
\graphicspath{ {./images/} }
\usepackage{amsmath,amsfonts}
\usepackage{algorithmic}
\usepackage{algorithm}
\usepackage{array}
\usepackage[caption=false,font=normalsize,labelfont=sf,textfont=sf]{subfig}
\usepackage{textcomp}
\usepackage{stfloats}
\usepackage{url}
\usepackage{verbatim}
\usepackage{graphicx}
\usepackage{cite}
\usepackage{multirow}
\usepackage{diagbox}
\usepackage{multicol}
\usepackage{makecell}
\usepackage[table]{xcolor}
\usepackage{xcolor}
\usepackage[utf8]{inputenc}
\usepackage{textgreek}

\title{DMVFC: Deep Learning Based Functionally Consistent Tractography Fiber Clustering Using Multimodal Diffusion MRI and Functional MRI}

\author{
  Bocheng Guo\footnotemark[1]\hspace{5pt}\footnotemark[2] \\
  \And
  Jin Wang\footnotemark[1]\hspace{5pt}\footnotemark[2] \\
  \And
  Yijie Li\footnotemark[2] \\
  \And
  Junyi Wang\footnotemark[2] \\
  \And
  Mingyu Gao\footnotemark[2] \\
  \And
  Puming Feng\footnotemark[2] \\
  \And
  Yuqian Chen\footnotemark[3] \\
  \And
  Jarrett Rushmore\footnotemark[4] \\
  \And
  Nikos Makris\footnotemark[3] \\
  \And
  Yogesh Rathi\footnotemark[3] \\
  \And
  Lauren J O'Donnell\footnotemark[5]\hspace{5pt}\footnotemark[3] \\
  \And
  Fan Zhang\footnotemark[6]\hspace{5pt}\footnotemark[5]\hspace{5pt}\footnotemark[2]
}
\begin{document}
\maketitle
\begin{abstract}
Tractography fiber clustering using diffusion MRI (dMRI) is a crucial method for white matter (WM) parcellation to enable analysis of brain’s structural connectivity in health and disease. Current fiber clustering strategies primarily use the fiber geometric characteristics (i.e., the spatial trajectories) to group similar fibers into clusters, while neglecting the functional and microstructural information of the fiber tracts. There is increasing evidence that neural activity in the WM can be measured using functional MRI (fMRI), providing potentially valuable multimodal information for fiber clustering to enhance its functional coherence. Furthermore, microstructural features such as fractional anisotropy (FA) can be computed from dMRI as additional information to ensure the anatomical coherence of the clusters. In this paper, we develop a novel deep learning fiber clustering framework, namely \textit{Deep Multi-view Fiber Clustering (DMVFC)}, which uses joint multi-modal dMRI and fMRI data to enable functionally consistent WM parcellation. DMVFC can effectively integrate the geometric and microstructural characteristics of the WM fibers with the fMRI BOLD signals along the fiber tracts. DMVFC includes two major components: (1) a multi-view pretraining module to compute embedding features from each source of information separately, including fiber geometry, microstructure measures, and functional signals, and (2) a collaborative fine-tuning module to simultaneously refine the differences of embeddings. In the experiments, we compare DMVFC with two state-of-the-art fiber clustering methods and demonstrate superior performance in achieving functionally meaningful and consistent WM parcellation results. 
\end{abstract}

\section{Introduction}
Diffusion magnetic resonance imaging (dMRI) tractography is a well-established neuroimaging technique that uniquely allows in vivo mapping of the brain's white matter (WM) connections at the macroscopic level \cite{basser2000vivo}. This technique has been widely used in the quantitative analysis of the brain's structural connectivity \cite{zhang2022quantitative}. Fiber clustering is a crucial strategy for WM parcellation to subdivide whole-brain tractography into geometrically similar and anatomically meaningful bundles \cite{garyfallidis2012quickbundles},  \cite{zhang2018anatomically}. Although many fiber clustering methods have been developed for parcellation of the WM tracts  \cite{zhang2022quantitative}, existing clustering methods share a common limitation that they do not explicitly capture the functional and microstructural implications of fiber clusters. To achieve clustering outcomes that are both anatomically and functionally meaningful, it is crucial to develop a framework that integrates both structural and functional modalities.

\subsection{Tractography Parcellation Using Fiber Clustering}
Tractography parcellation is essential for quantitative analysis of the brain’s structural connectivity. Fiber clustering is an important tractography parcellation strategy to organize streamlines into coherent groups based on their geometric similarities \cite{zhang2022quantitative}. Briefly, fiber clustering methods begin by computing pairwise similarity distances between fibers and subsequently group them using various clustering algorithms. Compared to other tractography parcellation methods, such as the cortical-parcellation-based methods that organize fibers by their termination points in gray matter areas, fiber clustering techniques yield greater inter-subject consistency and exhibit improved reproduction reliability \cite{zhang2019test,zhang2017comparison}. This methodological robustness has enabled deeper investigation into the organization of WM throughout different stages of life and disease conditions \cite{zhang2018anatomically}. Ji et al applied a novel atlas-based parcellation approach to cluster and analyze superficial WM U-fibers in schizophrenia and bipolar disorder \cite{ji2019increased}. Automatically annotated fibre clustering (AAFC)\cite{feng2020local} facilitates detailed analysis of WM tract microstructure, enabling effective differentiation between Parkinson’s disease, Scans Without Evidence of Dopaminergic Deficit (SWEDD), and healthy controls.

Many approaches have been proposed for clustering tractography data. QuickBundles, for instance, employs a computationally efficient clustering algorithm based on the minimum average direct-flip distance to rapidly cluster streamlines for individual subjects \cite{garyfallidis2012quickbundles}. Beyond individual-level clustering, groupwise methods have been developed to create population-based tractography atlases. One example is WhiteMatterAnalysis (WMA), which applies spectral clustering using pairwise distances between closest fiber points to segment fibers across multiple subjects \cite{zhang2018anatomically}, \cite{o2007automatic}. FFClust is another, a high-performance algorithm that is designed to utilize graph modeling, facilitating the identification of compact and anatomically meaningful WM bundles in both individual and population-level analyses \cite{vazquez2020ffclust}. Román et al employ a two-level hierarchical clustering approach by first segmenting fiber centroids into ROI-based subgroups for initial clustering, followed by a second clustering stage to identify the most reproducible short association fiber bundles across subjects\cite{roman2022superficial}.  These techniques help capture consistent anatomical patterns across individuals and are critical in studies of population-level WM organization. Recent advances in deep learning have enabled enhanced effective and efficient parcellation of WM fiber tracts through innovative approaches that leverage graph network\cite{liu2019deepbundle}, self-supervised learning \cite{chen2023deep}, and contrastive learning techniques \cite{xue2022supwma}.

These methods offer distinct frameworks that improve anatomical accuracy and computational speed, addressing challenges in brain WM microstructure analysis. However, existing fiber clustering methods primarily rely on geometric features and have yet to integrate fMRI data to ensure that fiber clusters exhibit functional consistency. 

\subsection{Functional and Microstructural Information Along WM Fiber Tracts}
Functional magnetic resonance imaging (fMRI), which utilizes blood oxygen level-dependent (BOLD) contrast \cite{garyfallidis2012quickbundles}, is a well-established method for assessing functional activity in the brain's gray matter (GM). It has long been assumed that BOLD signals mainly arise from postsynaptic potentials in GM, where such activity is virtually absent in WM, leading to ongoing debate about the functional significance of BOLD responses observed in WM. However, recent research has shown that BOLD signals in WM consistently appear during visual and motor tasks, displaying patterns that closely resemble those found in GM \cite{courtemanche2018detecting}, though with lower amplitudes and a slight time delay \cite{fraser2012white}. Notably, the resting-state frequency spectrum of WM BOLD signals has been found to closely match the spectral profile characteristic of GM \cite{li2021power}, \cite{huang2018voxel}. Furthermore, stable, long-range functional networks can be reliably detected within WM at the voxel level \cite{peer2017evidence}, and these networks are well correlated with those in GM, which indicates functional homogeneities of WM BOLD signals within fiber tracts \cite{ding2018detection}. Therefore, a promising solution for the construction of brain functional architecture is to exploit the functional signal transmission along WM fibers, which will take advantage of the abundant multimodal information, including the spatial distribution and temporal sequence of BOLD signals.

Furthermore, from the dMRI data itself, many quantitative measures can be computed to reflect the underlying tissue microstructure properties of the WM. For example, fractional anisotropy (FA) is a neuroimaging index of microstructural WM integrity. This measure is sensitive to a variety of microstructural features. FA may reflect the coherence of membranes and restrictions \cite{kochunov2012fractional}. However, to the best of our knowledge, no study has incorporated FA into fiber clustering yet.  Moreover, FA is highly sensitive to the fractional composition of tissue that includes myelinated neurons \cite{schilling2022anomalous}. As FA reflects the morphological characteristics of fibers, it provides valuable insights for clustering fibers with similar structural properties. This suggests that incorporating FA could be particularly advantageous for achieving more anatomically coherent clustering outcomes.

\subsection{Unsupervised Deep Feature Learning and Multi-view Clustering}
Deep neural networks have demonstrated superior performance in various computational neuroimaging tasks such as registration \cite{zhang2021deep},  imaging \cite{huang2021brain}, and segmentation \cite{wang2022accurate}. In particular, deep-learning-based clustering has been the subject of extensive study as an unsupervised learning task \cite{ren2024deep}, which can be a powerful tool in fiber clustering analysis. One straightforward approach of  unsupervised deep clustering is to extract feature embeddings with neural networks and then perform clustering on these embeddings to form clusters. The learned embeddings are high-level representations of input data and have been demonstrated to be informative for downstream tasks. For fiber clustering, studies have been proposed to use self-supervised learning. CINTA \cite{legarreta2021filtering}, an unsupervised autoencoder-based method for tractography clustering, avoids distance thresholding and achieves anatomically coherent results with linear time complexity. Deep Fiber Clustering (DFC) \cite{chen2023deep} framework utilizes self-supervised learning to derive meaningful embeddings for clustering WM fibers.\cite{xie2022self}

A promising direction is using multi-view clustering for effective WM parcellation to combine fMRI information with dMRI information. In recent years, benefiting from the ability to well exploit the underlying information embedded in the data from different views in unsupervised clustering, multi-view clustering (MVC) has been increasingly researched and applied to multiple scenarios. Multi-view clustering methods like \cite{cui2023novel}, \cite{xu2021deep} have gained better performance on vision multi-view datasets than single-view datasets. However, the computer vision tasks are based on the different views of the same modality data. On fibers, fMRI and dMRI yield complementary insights, where fMRI reveals functional activity along the pathways, and dMRI maps geometric trajectories and structural information. It inspires us to utilize the attribute of integrating complementary information from multi-view clustering. Although fMRI provides valuable insights into membrane coherence and microstructural constraints, to date, no studies have incorporated fMRI as a fusion mechanism within multi-view clustering frameworks for WM fibers.

\subsection{Contributions}
In this work, we present a new multimodal fiber clustering framework, Deep Multi-view Fiber Clustering (DMVFC), that combines multiple types of data to improve cluster connectivity correlation and detail of fiber clustering analysis. In contrast to previous studies primarily focused on dMRI \cite{garyfallidis2012quickbundles}, \cite{roman2022superficial}, \cite{legarreta2021filtering}, we put forth a more comprehensive self-supervised multi-view deep learning framework with additional information as a clustering constraint. We utilize multimodal data to perform WM clustering efficiently, achieving fiber clustering with functional and microstructural consistency. The final results of our experiments evaluate the feasibility of our method, and the comparison with previous methods highlights the reliability and advantages of our approach. Source code is available at: https://github.com/GBCWORLDWALKER/DMVFC. 

We have three main contributions. First, we apply deep multi-view clustering for fiber clustering, providing a novel approach to analyzing neural pathways. Second, our method uniquely incorporates dMRI data and fMRI data jointly into deep fiber clustering, leading to enhanced performance in clustering the brain's structural and functional connectivity. Third, by integrating different data modalities, our work provides more insights and establishes a new framework for fiber clustering techniques. The preliminary version of this work, referred to as DMVFC$_{conf}$, was published in ISBI 2025 \cite{wang2025novel}. In this work, we extend our previous work by 1) optimizing the fine-tuning process by using multi-modal based initialization, 2) adding FA information to further improve cluster anatomical coherence, and 3) a comprehensive evaluation on additional anatomical fiber tracts. 

\begin{figure*}[!t]
\centering
\includegraphics[width=0.8\linewidth]{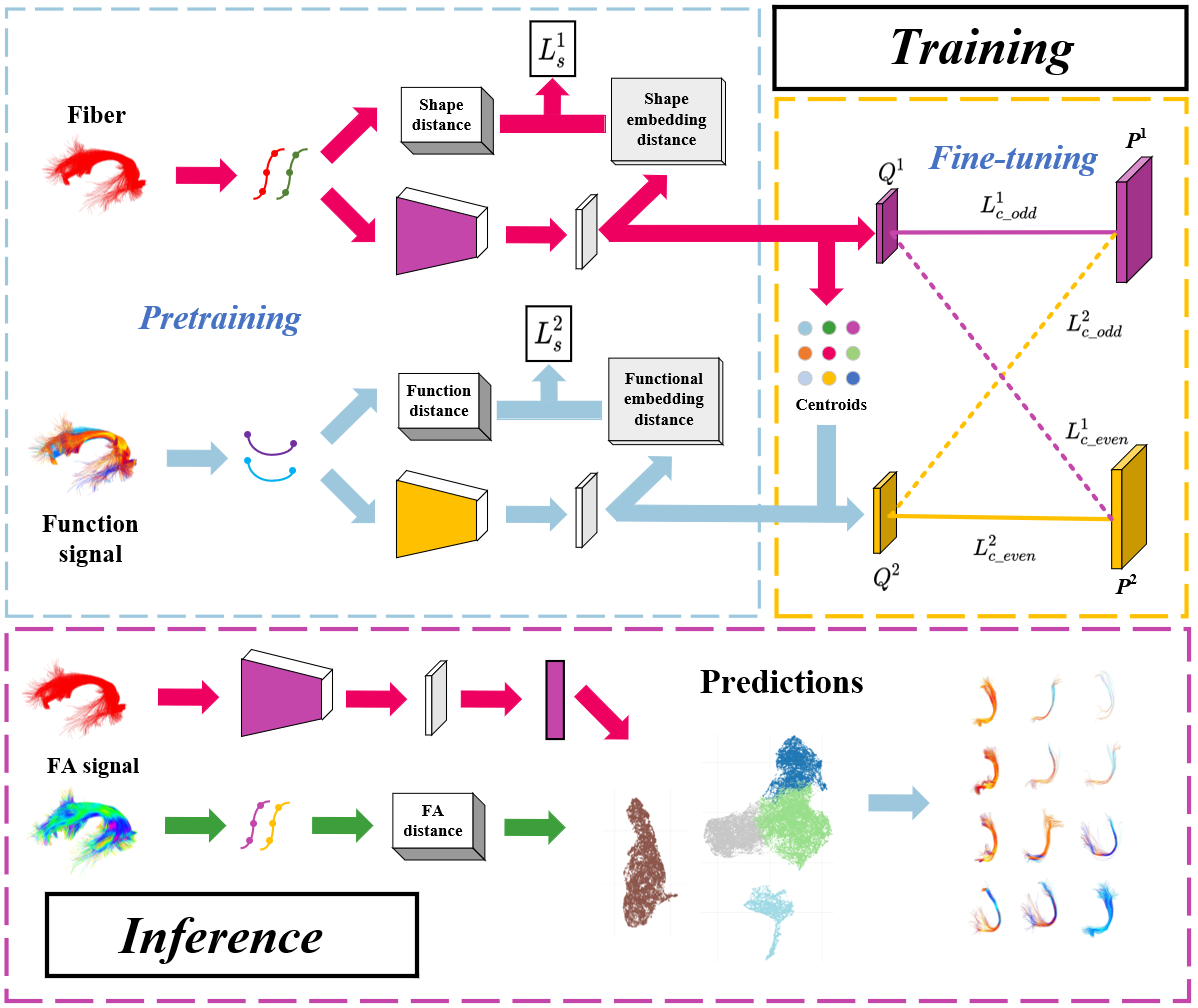}%
\hfil
\caption{Method Overview. 
The training stage consists of pretraining and tuning processes. During the pretraining stage, two parallel feature extraction models are trained to compute embeddings from fiber geometric information and brain functional signals, respectively. During the fine-tuning stage, the pretrained embeddings are optimized to ensure that the clustering outcomes integrate both geometric and functional information simultaneously. During the inference stage, fiber geometry information and FA data are incorporated to optimize clustering results. During this phase, an input fiber is assigned to the cluster with the highest soft label assignment probability, indicating the specific cluster to which the fiber belongs.}
\label{fig:fig1}
\end{figure*}

\section{METHODS}
The overall workflow of DMVFC is shown in Figure \ref{fig:fig1}. There are two main training stages in DMVFC: (1) multi-view pretraining and (2) collaborative fine-tuning. In the first stage, the multi-view pretraining module computes two kinds of embeddings from fiber geometric information and brain functional signals (Section \ref{sec:pretraining}). In the second stage, the collaborative fine-tuning module refines the pretrained embeddings to ensure that the clustering outcomes integrate both dMRI and fMRI information simultaneously (Section \ref{sec:Collaborative Fine-tuning}). During the inference stage (Section \ref{sec:inference stage}), fractional anisotropy (FA) is incorporated as supplementary information to enhance the clustering process.

\subsection{Multi-view Pre-training for Embedding Computation}
\label{sec:pretraining}
In multi-view clustering analysis, the term “views” refers to different representations or perspectives of the same dataset, enabling the clustering algorithm to exploit complementary information from each view to enhance clustering accuracy and robustness \cite{chen2020multi}. In our work, the multi-view pre-training module consists of two parallel embedding extraction networks designed to generate feature embeddings from two distinct views, i.e., the fiber geometry information and brain functional signals, respectively. The two models are based on the popular dynamic graph convolutional neural network (DGCNN) [32], which is specifically designed for processing graph-structured data such as point clouds. DGCNN has been shown to be successful for fiber clustering in the Deep Fiber Clustering (DFC) framework \cite{chen2023deep}, where it leverages geometric similarity between fibers to derive embedding features and subsequently refines the embedding space based on clustering outcomes. In the proposed DMVFC, we extend this approach to a multi-view model that incorporates both fiber geometry and functional signals.

\subsubsection{Network inputs}
\label{sec:Nestwork inputs}
To represent geometric information, fibers are modeled as point clouds that are defined by the spatial coordinates of fiber points along the trajectories. Specifically, for a given fiber $i$, its geometric input is denoted as $\mathbf{x}_i^1\in \mathbb{R}^{n_p \times 3}$, where $n_p$ corresponds to the number of evenly sampled points along the fiber, represented in the spatial Right-Anterior-Superior (RAS) coordinate system.

For capturing functional signals, fibers are represented as point clouds defined by the BOLD signals at the endpoints. The fiber endpoints are located in the grey matter (except for tracts such as the CST, where one endpoint is within the white matter brainstem region).To improve cross-subject comparability and measurement robustness, we use the BOLD signals at the two endpoints near the cortex as the functional representation: endpoints lie near the cortex, provide a higher signal-to-noise ratio, and facilitate parcellation-based alignment. By contrast, correlations of along-tract white-matter BOLD can be susceptible to local confounds—e.g., small vessel/venous signals, crossing-fiber interference and partial-volume effects, field inhomogeneity, and regional variability in the hemodynamic response function—leading to larger phase/latency variability and noise and thus hampering cross-subject consistency assessments. Each endpoint contains 1,200 time points (see Section \ref{sec:III A}), which were randomly downsampled to 600 to enable more efficient CPU processing. As a result, for a fiber $i$, its functional signal input is formed as $\mathbf{x}_i^2\in \mathbb{R}^{2\times 600}$.

\subsubsection{Network training}

To obtain geometric and functional signal embeddings that effectively distinguish each fiber and improve clustering performance, we adopt the design principles of the DFC method \cite{chen2023deep}. DFC is a self-supervised learning-based method that processes pairs of fibers as inputs and employs fiber distance measures as pseudo-labels to train latent embedding representations for each fiber via DGCNN.

In contrast to DFC’s single-model design, our method employs two parallel DGCNN networks to process the geometric and functional inputs represented as:
\begin{equation}
\left\{ (\mathbf{x}_i^v, \mathbf{y}_i^v) \mid \mathbf{x}_i^v, \mathbf{y}_i^v \in \mathbb{R}^{n_{p}\times C}\right\}_{i=1}^N
\end{equation}

Specifically, consider the dataset where $(\mathbf{x}_i^v,\mathbf{y}_i^v)$  denotes an input pair fed into its respective DGCNN network. Here, $v$ indicates which view a particular input belongs to, $N$ represents the total number of fibers, $n_p$ represents the number of points per fiber, and $C$ corresponds to the number of channels for each data type. For geometric input ($v=1$), the pseudo-label $\mathbf{s}_i^1$ is derived from the minimum average direct-flip distance, a metric commonly used in WM fiber clustering \cite{garyfallidis2012quickbundles}, \cite{zhang2018anatomically}. In contrast, for functional input ($v=2$), pseudo-label $\mathbf{s}_i^2$ is calculated using the Pearson correlation of fMRI signals.

Finally, to ensure that the distance between embeddings aligns with the similarity of data pairs, our study employs the following loss function:

\begin{equation}
L_S=\sum_{i=1}^N||d(\mathbf{x}_i^v,\mathbf{y}_i^v)-\mathbf{s}_i^v||_2^2
\label{eq:2}
\end{equation}
where $\mathbf{s}_i^v$ represents the label of input pairs, d(•) denotes the Euclidean distance between the learned deep embeddings $f_\theta^v(\mathbf{x}_i^v)$.

\subsection{Collaborative Fine-tuning}
\label{sec:Collaborative Fine-tuning}
Since geometric and functional embeddings of fibers are computed independently, integrating the complementary information is crucial to achieve consistent clustering results across both modalities. To achieve this, we draw inspiration from the deep embedded multi-view clustering with collaborative training (DEMVC) method \cite{xu2021deep}, which leverages embeddings from different views to mutually guide the fine-tuning of each other. In this case, this collaborative algorithm allows each embedding to assimilate complementary features from other views through the fine-tuning process.

In our collaborative fine-tuning stage, the overall loss is represented as a weighted sum of $L_s$ (see Eq. \ref{eq:2}) to preserve embedding similarity between different fibers, along with $L_c$, which serves as an alternative collaborative loss function (refer to Eq. \ref{eq:3} and Eq. \ref{eq:4}), applied alternately during odd and even training epochs as shown in the following equations. This alternating strategy helps the model learn from different perspectives and prevents it from overly relying on specific patterns.

\begin{equation}
\label{eq:3}
L_{c\underline{~}odd}^{v}=KL(P^1||Q^v)=\sum_{i=1}^N\sum_{j=1}^Kp_{ij}^1\mathrm{log}\frac{p_{ij}^1}{q_{ij}^v}
\end{equation}

\begin{equation}
\label{eq:4}
L_{c\underline{~}even}^{v}=KL(P^2||Q^v)=\sum_{i=1}^N\sum_{j=1}^Kp_{ij}^2\mathrm{log}\frac{p_{ij}^2}{q_{ij}^v}
\end{equation}
where $Q^v$ denotes the soft label of all fibers across different views (with $v=1$ representing the geometric view and $v=2$ denoting the functional view). $q_{ij}^v$ represents the probability that one input fiber belongs to a given cluster $j$, which is defined by the t-student distribution:

\begin{equation}
q_{ij}^v=\frac{(1+||z^v_i-\mu^v_j||^2)^{-1}}{\sum_{j'}(1+||z^v_i-\mu^v_{j'}||^2)^{-1}}
\end{equation}
where $z^v_i$ is the embedding of an individual fiber which has index $i$ and ${\mu}^v_j$ is the centroid of cluster $j$, This employs the Student’s t-distribution to convert the distance between individual fiber and cluster centroids into probabilities for soft label assignment, leveraging the heavy-tailed property of the t-distribution to ensure a smooth and robust probability distribution. $P^v$ is the target distribution and $p_{ij}^v$ is defined as:

\begin{equation}
p_{ij}^v = \frac{(q_{ij}^v)^2 / \sum_i q_{ij}^v}{\sum_j \left( (q_{ij}^v)^2 / \sum_i q_{ij}^v \right)}
\end{equation}

It is notable that the initialization of the centroids of clusters at the very beginning of fine-tuning process in both geometric information view and functional information view are crucial for the final results of optimization of the model. We calculate the centroids of geometric information embeddings first, and then use the fiber indexes of centroids in geometric information view as the centroids of the functional information embedding. The impact of the initialization strategy is further examined in the ablation study presented in Section \ref{sec:III C}.  

\begin{figure*}
\centering
\includegraphics[width=5.5in]{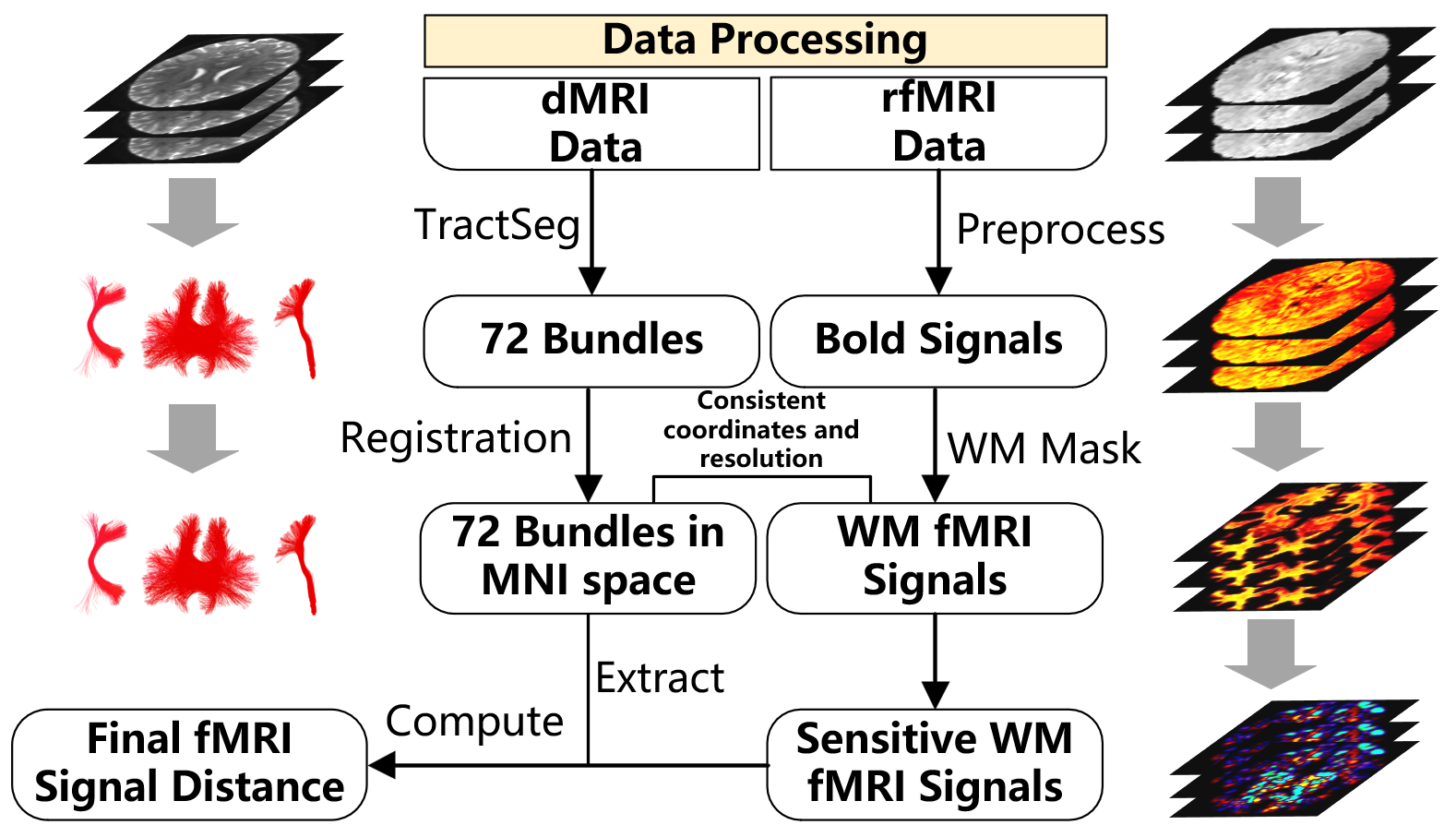}%
\hfil
\caption{Flowchart of data preprocessing}
\label{fig:fig2}
\end{figure*}

During the training process, the clustering losses are applied alternately in odd and even epochs. The fine-tuning Loss is defined as :

\begin{equation}
L_f=L_s+\gamma L_c
\end{equation}
where $\gamma$ is empirically set to be 0.1, $L_c$ is defined as $L_{c\underline{~} odd}$ during odd epochs and $L_c = L_{c \underline{~} even}$ during even epochs.

\subsection{Inference Stage}
\label{sec:inference stage}
During the inference stage, for a new subject, we first compute the FA similarity between individual fibers. In parallel, the geometric input (as introduced in Section \ref{sec:Nestwork inputs}) is constructed and fed into the network to obtain the corresponding embeddings. The Manhattan distance is employed to quantify the FA similarity between fibers. This similarity value is then scaled by a factor of 30 and added to the geometric distance produced by the model to compute the final distance. This final distance is subsequently used to estimate the probability of each input belonging to a given cluster, as described in Eq. 5 (Section \ref{sec:Collaborative Fine-tuning}). The final cluster assignment for each fiber is determined by selecting the cluster with the highest probability. Notably, fMRI data are not needed at the inference stage. The model operating in the geometric information view is trained with guidance from functional data, but performs prediction independently of it.

\section{EXPERIMENTS AND RESULTS}
\label{sec:III A}
\subsection{Data Acquisition and Preprocessing}
In this study, we utilize well-preprocessed dMRI and rfMRI data from the unrelated 100 subjects in the Human Connectome Project Young Adult (HCP-YA) \cite{van2013wu}. For each participant, one hour of whole-brain rsfMRI data were collected using a 3 Tesla scanner, with a spatial resolution of 2×2×2 mm$^3$, repetition Time (TR) of 720 ms, echo time (TE) of 33.1 ms, dMRI data were obtained with a spatial resolution of 1.25×1.25×1.25 mm, TR of 5520 ms, and TE of 89.5 ms.

The overall data processing workflow is illustrated in Figure \ref{fig:fig2}. Diffusion MRI data were processed using the widely adopted tract segmentation tool TractSeg \cite{wasserthal2018tractseg} to delineate anatomical fiber bundles. In our experiment, all 72 bundles were analysed. The rsfMRI data are preprocessed through the HCP fMRI minimal pipeline \cite{glasser2013minimal}. Then they were cleaned of spatially specific structured noise (ICA-FIX) \cite{griffanti2014ica} and precisely aligned across subjects by multimodal cortical surface registration (MSMAll) \cite{glasser2013minimal}. To prevent signal contamination and improve the signal-to-noise ratio and fMRI sensitivity in WM, we carry out spatial smoothing (4 mm full-width half-maximum, isotropic) for GM and WM separately. These preprocessing steps are done using the SPM12 toolbox \cite{ashburner2014spm12}. Finally, the WM signals were band-pass filtered within the 0.01 to 0.08 Hz frequency range.

\subsection{Implementation}
The entire dataset was divided into 80 training samples and 20 testing samples. During the pretraining phase, the model was trained for 450 epochs with an initial learning rate of $3\times10^{-3}$, which decayed by a factor of 0.1 every 200 epochs.  This learning rate decay strategy helps the model converge more effectively by allowing larger updates in the early training stages to accelerate learning, while gradually reducing the learning rate for more stable updates later in training. In the subsequent fine-tuning phase, training was conducted for 20 epochs using a learning rate of $1\times10^{-5}$. A batch size of 1024 was employed, and optimization was performed using the Adam algorithm \cite{kingma2014adam}. All deep learning models were implemented with PyTorch (v2.4.1) and Python (3.10). Training and inference were carried out on a single NVIDIA 3090 GPU. The training time for each bundle is approximately 6 hours, and the inference time is about 15 minutes. Visualization of results was conducted using 3D Slicer via SlicerDMRI \cite{zhang2020slicerdMRI} \cite{norton2017slicerdMRI}. The fiber bundles selected for quantitative evaluation in Tables I-III and Figure 3 were randomly chosen to demonstrate the effectiveness of the paper’s core method.

To ensure fast and efficient processing of the large number of fiber samples during model training and inference, fibers are downsampled to 25 points, as this number provides good performance with relatively low computational costs. 

\begin{table*}[t]
\renewcommand{\arraystretch}{1.7}
\normalsize
\caption{Average GM correlation and $\alpha$ in different bundles under different model.}
\label{tab:tab1}
\centering
\setlength{\tabcolsep}{3.9mm}
\begin{tabular}{ccccccccccc}%

\toprule
\multirow{2}{*}{Bundle}&\multicolumn{2}{c}{Quick Bundle}&\multicolumn{2}{c}{\makecell{{Riemannian}\\{Framework}}}&\multicolumn{2}{c}{DFC}&\multicolumn{2}{c}{DMVFC$_{conf}$}&\multicolumn{2}{c}{DMVFC}\\
\cmidrule{2-11}
&Corr&$\alpha$&Corr&$\alpha$&Corr&$\alpha$&Corr&$\alpha$&Corr&$\alpha$\\
\midrule
ATR&0.266&4.991&0.322&4.762&0.308&4.544&0.318&4.582&\textbf{0.347}&\textbf{4.521}\\
CG&0.358&5.603&0.360&5.610&0.372&5.570&0.373&5.565&\textbf{0.378}&\textbf{5.561}\\
CST&0.251&5.124&0.291&4.723&0.311&4.288&0.314&4.289&\textbf{0.326}&\textbf{4.281}\\
FPT&0.339&4.931&0.392&4.914&0.374&4.890&0.379&5.068&\textbf{0.394}&\textbf{4.886}\\
ICP&0.253&3.968&0.272&3.991&0.315&4.003&0.332&4.415&\textbf{0.341}&\textbf{3.914}\\
IFO&0.275&5.605&0.311&5.599&0.319&5.573&0.318&5.576&\textbf{0.322}&\textbf{5.571}\\
POPT&0.234&5.264&0.285&5.261&0.275&5.242&0.293&5.380&\textbf{0.294}&\textbf{5.231}\\
SCP&0.251&5.124&0.291&4.723&0.349&4.286&0.354&4.304&\textbf{0.367}&\textbf{4.235}\\
ST\_FO&0.264&4.315&0.364&4.002&0.469&3.715&0.470&3.747&\textbf{0.478}&\textbf{3.711}\\
ST\_PREC&0.307&4.435&0.377&4.422&0.381&4.407&0.381&4.412&\textbf{0.384}&\textbf{4.405}\\
ST\_PAR&0.301&5.423&0.340&5.417&0.333&5.457&0.338&5.416&\textbf{0.348}&\textbf{5.398}\\
STR&0.532&3.193&0.533&3.188&0.531&3.159&0.532&3.180&\textbf{0.536}&\textbf{3.004}\\
\bottomrule
\label{table_main}
\end{tabular}
\end{table*}

\subsection{Experimental Results}
\label{sec:III C}
\subsubsection{Model  performance comparison with state-of-the-art methods}
To evaluate our clustering results, we compare our method with three state-of-the-art fiber clustering methods: QuickBundles (QB) \cite{garyfallidis2012quickbundles}, a Riemannian Framework for Structurally Curated Functional Clustering of Brain WM Fibers \cite{zhao2023riemannian}, and Deep Fiber Clustering (DFC) \cite{chen2023deep}. In brief, QB is an efficient fiber clustering technique that groups similar fibers based on extracted features and spatial proximity. The Riemannian Framework is an advanced method for functional fiber clustering using WM BOLD signals along fibers. DFC is a deep learning-based method on which our method is built (see Section \ref{sec:pretraining}). 

We use the following two evaluation metrics. First, as the goal of functional fiber clustering is to ensure that fibers within a cluster exhibit functional homogeneity, we compute the Pearson correlation of fMRI signals at the endpoints of fibers within each cluster \cite{zhao2023riemannian}. A higher correlation indicates a stronger functional correlation across fibers within the cluster. In addition, to assess the geometric similarity of the fibers within a cluster, we compute the $\alpha$ measure, which is defined as the average pair-wise distances between all fibers within each cluster \cite{vazquez2020ffclust}. The value of α captures the coherence of the streamlines within clusters. A lower value of α indicates better coherence and improved clustering performance.

The average fMRI signal correlation and the $\alpha$ measure for clusters of the 12 bundles of interest are shown in Table \ref{table_main}. For all the bundles shown in the table, we calculate the average metrics for the bilateral bundles. Our method in general outperforms the other methods in terms of functional performance and maintains the lowest $\alpha$ value. These data demonstrate that our clustering technique not only ensures high functional relevance but also maintains strong geometric consistency.
\begin{figure}[H]

    \centering
    \includegraphics[width=\linewidth]{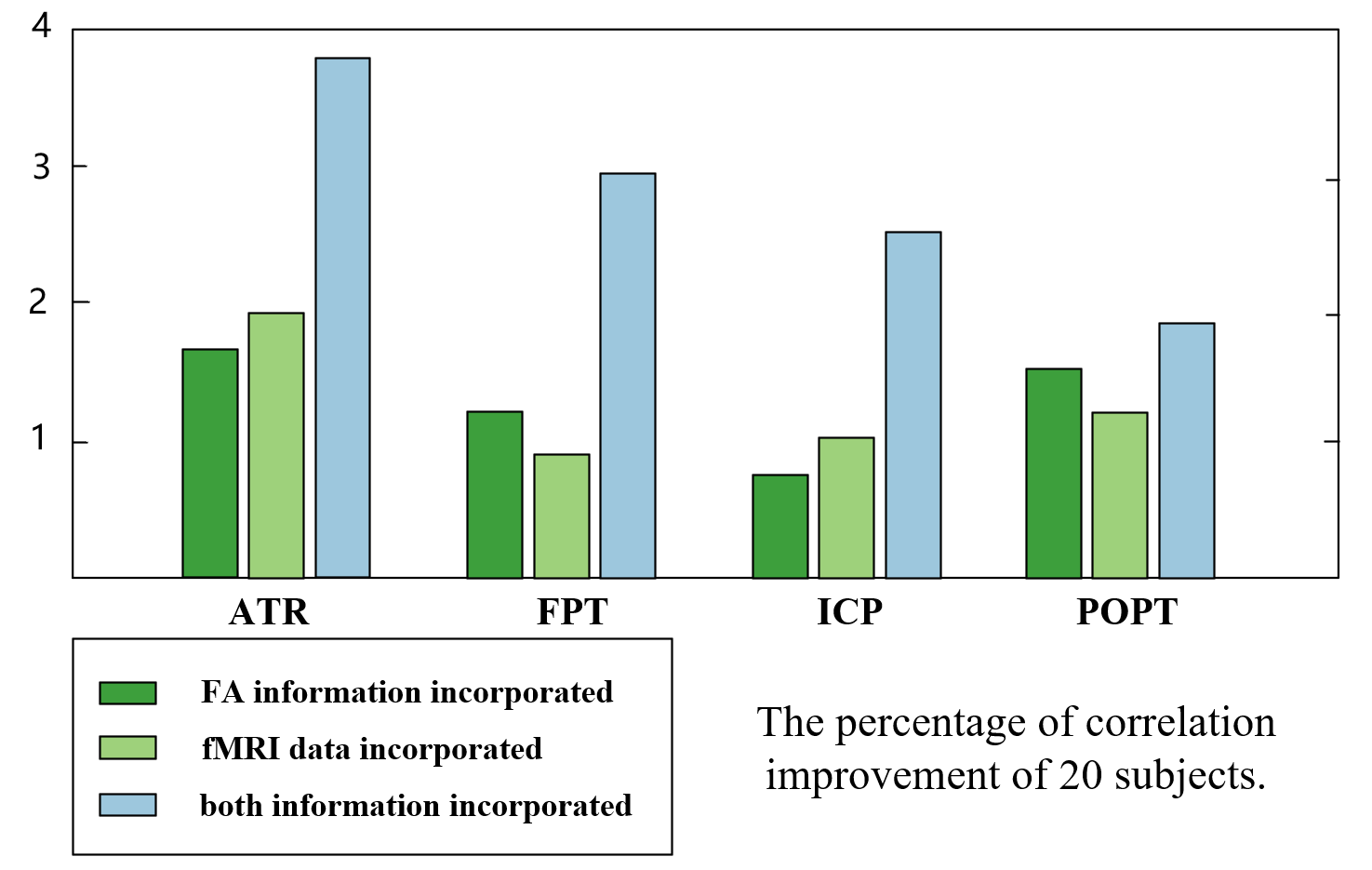}
    \caption{Impact of incorporating fMRI and FA information on model performance}
    \label{fmri_fa}
\end{figure}
\subsubsection{Ablation study of embedding initialization before fine-tuning}
As discussed in Section \ref{sec:Collaborative Fine-tuning}, the initialization of the clustering embeddings plays a critical role in the subsequent fine-tuning process. To systematically evaluate its impact, we perform an ablation study designed to analyze the influence of different initialization strategies on the final model's performance. This investigation aims to provide empirical insights into how initialization affects model representation quality. Our method utilizes centroids of geometric information view to initialize the centroids of functional information view. In the ablation study, individual initialization method without explicit initialization relies on performing initialization within their respective embedding spaces by using centroids derived from each individual view. All other parameters and experimental conditions are held constant to isolate the effect of initialization strategy on model performance. Following this initialization approach, both model configurations undergo fine-tuning for 20 epochs. The metrics fMRI endpoints correlation and $\alpha$ are calculated on the test datasets. This design enables a clear assessment of how omitting the specific initialization impacts the fine-tuning process and final outcomes. As shown in Table \ref{table_initial}, our method achieves higher fMRI correlation and less $\alpha$. It indicates that the method of utilizing geometric information embedding centroids to initialize functional information view has a positive effect on subsequent fine-tuning.

\subsubsection{Ablation study of incorporation of fMRI and FA information}
To evaluate the impact of incorporatin fMRI and FA information on model performance, we conducted a comparative analysis involving three distinct model configurations relative to a baseline model: (1) the proposed model incorporating both FA and fMRI data, (2) the baseline model enhanced solely with fMRI data, and (3) the baseline model augmented exclusively with FA information. The fMRI-enhanced model was implemented using the same multi-view collaborative clustering framework as the proposed model, with the FA component removed to isolate the effect of fMRI data. Conversely, the FA-only model is based on the DFC \cite{chen2023deep} framework, incorporating FA information exclusively during the inference stage, as detailed in Section \ref{sec:inference stage}. These results are summarized in Figure \ref{fmri_fa}, where DFC\cite{chen2023deep} serves as the reference baseline, and improvements in fMRI correlation are depicted via bar plots. The findings clearly indicate that integrating FA information leads to consistent and significant enhancements in model performance across all configurations compared to the baseline. This suggests that FA provides valuable complementary information, thereby improving the robustness and accuracy of brain functional modeling. Moreover, the proposed model integrating both fMRI and FA information demonstrates superior improvement compared to models incorporating each modality individually. Notably, in certain fiber bundles, the performance improvement exceeds the additive effect of combining the separate improvements observed from fMRI and FA alone, indicating a synergistic interaction between the two data types.

\begin{table}[!t]
\renewcommand{\arraystretch}{1.5}
\normalsize
\caption{Ablation study of centroid initialization in the embedding space. fMRI correlation and alpha metrics for the final results are reported for models using different initialization strategies before fine-tuning.}
\centering
\setlength{\tabcolsep}{1mm}{
\begin{tabular}{lcccccc}
\toprule
\diagbox{Bundle}{Method} & & ATR & CG & CST & FPT & ICP \\
\midrule
\multirow{2}{*}{\makecell[c]{\textbf{\hspace{1em}DMVFC}}} & Corr & \textbf{0.347} & \textbf{0.378} & \textbf{0.326} & \textbf{0.394} & \textbf{0.341} \\
& $\alpha$ & \textbf{4.521} & \textbf{5.561} & \textbf{4.281} & \textbf{4.886} & \textbf{3.914} \\
\multirow{2}{*}{\makecell[c]{\textbf{Individual} \\ \textbf{Initialization}}} & Corr & 0.322 & 0.377 & 0.316 & 0.390 & 0.336 \\
& $\alpha$ & 4.554 & 5.564 & 4.288 & 4.994 & 4.143 \\
\bottomrule
\label{table_initial}
\end{tabular}
}
\end{table}

\begin{table}[!t]
\renewcommand{\arraystretch}{2}
\caption{Hausdorff Distance between corresponding functional pathways, fibers in clusters, and fibers in bundles (MD denotes Mean Hausdorff Distance)}
\centering

\setlength{\tabcolsep}{1mm}{
\begin{tabular}{cccccc}
\toprule
\multirow{2}{*}[-2pt]{Bundle} & \multicolumn{3}{c}{Pairwise MD of pathways} & \multirow{2}{*}[-2pt]{MD in clusters} & \multirow{2}{*}[-2pt]{MD in bundles} \\
\cmidrule{2-4}
& sub1-sub2 & sub1-sub3 & sub2-sub3 & & \\ 
\midrule
ATR    & 6.43 & 5.27 & 6.58 & 7.35 & 13.52 \\
CC\_2  & 9.53 & 8.62 & 8.33 & 10.40 & 21.31 \\ 
CG     & 5.12 & 6.52 & 6.81 & 8.91 & 18.07 \\
CST    & 4.14 & 5.11 & 4.77 & 7.32 & 12.47 \\
FPT    & 5.24 & 7.02 & 6.23 & 8.29 & 17.14 \\
ICP    & 4.81 & 3.76 & 4.22 & 5.81 & 13.38 \\
POPT   & 7.36 & 7.56 & 7.27 & 8.93 & 18.77 \\
\bottomrule
\label{consistency}
\end{tabular} 
}
\end{table} 
\subsubsection{Experiments on clustering consistency across subjects}
To further quantitatively assess clustering consistency among subjects, a representative pathway is identified for each cluster. This pathway is defined as the single fiber that exhibits the highest total functional correlation with all other fibers within the same cluster. The correlation is calculated based on the BOLD signals at the fibers' endpoints, making this representative pathway the 'functional centroid' that best captures the cluster’s shared functional characteristics. Distances are calculated between the corresponding representative pathways for different subjects. These values are then contrasted with the average distance within clusters and the mean distance throughout the entire bundle. The Hausdorff distance serves as a tool to gauge the uniformity of spatial arrangements in three-dimensional space, and Table \ref{consistency} presents these results. Specifically, the Hausdorff distance between fibers acts as a measure of proximity for the point collections that fibers traverse; it is a frequently employed structural metric in the context of fiber clustering. Moreover, the data shows that the mean intra-cluster distance is consistently smaller than the mean distance across the entire bundle, indicating that the resulting clusters are structurally compact and well-defined. Furthermore, the analysis reveals that the pairwise distances between representative pathways from different subjects are significantly smaller than the average distances within the clusters themselves. This finding demonstrates a high degree of consistency for the identified pathways across the subject pool, which shows the effectiveness of our method.

\begin{figure*}[!t]

\centering
\includegraphics[width=6in]{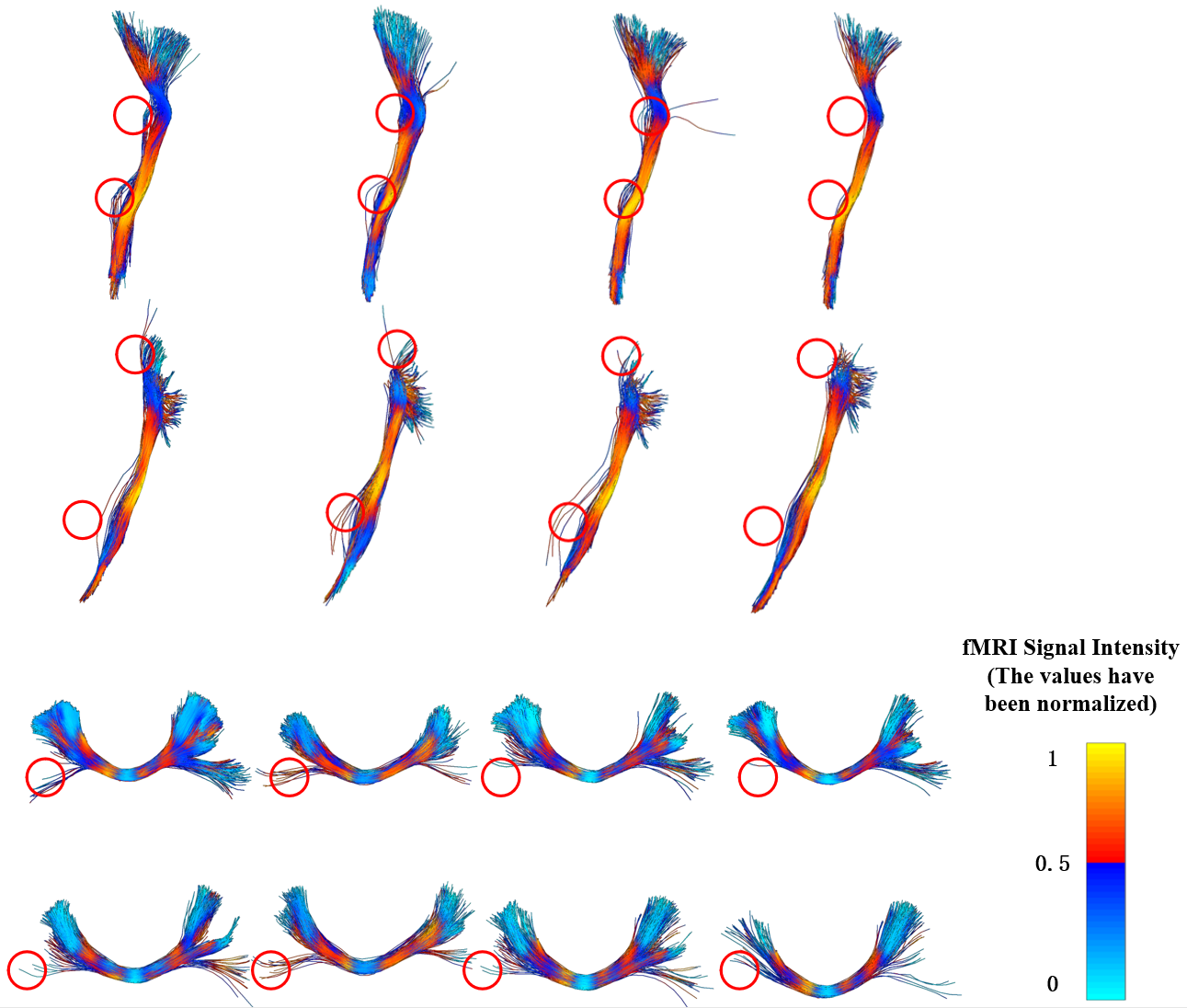}%
\hfil
\caption{Visualization of clustering results for four different methods, with clusters colored according to the strength of fMRI signals. Each column displays the corresponding cluster from three different clustering methods. Differences between clusters are highlighted by red circles. Note that these are selected example clusters for visualization purposes from all processed bundles, not exhaustive representations.}
\label{slicer_vis}
\end{figure*}

\begin{figure*}[!t]
\centering
\includegraphics[width=0.8\linewidth]{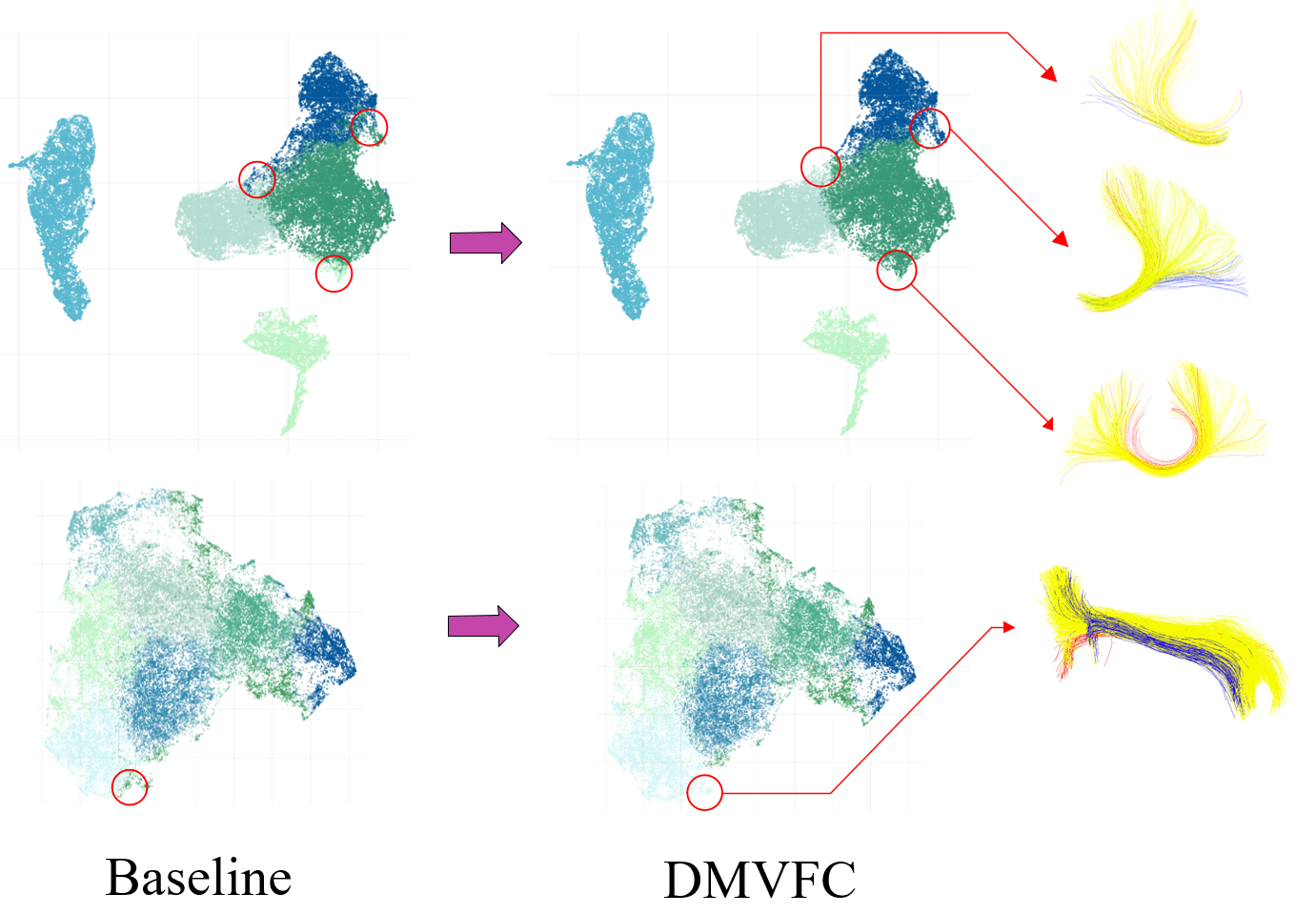}%
\hfil
\caption{Illustration of the geometric embedding using UMAP, with the baseline DFC on the left and our proposed framework in the middle, with the visualization of tractography of the corresponding bundle. Key differences are highlighted by the red circles. The blue streamlines represent those unique fiber to our method's cluster, while red streamlines indicate those exclusive to the baseline. Yellow streamlines denote the overlap, appearing in both our cluster and the baseline cluster. From all clustered white matter tracts, we select the CC\_5 (top) and SLF\_III (bottom) for demonstration.}
\label{embedding}
\end{figure*}

\subsubsection{Visualization of fiber clustering results}
To further demonstrate that our clustering results have functional homogeneity, we visualize the fMRI signals on several example clusters in Figure \ref{slicer_vis}. The colors in the clusters displayed in the figure correspond to the strength of the fMRI signals. Each column in the figure represents clusters that are spatially closest to one another within the same bundle, as identified by four different clustering methods. The detailed differences between the clusters, highlighted by red circles, illustrate the superior performance of our method in the removal of outliers and the clustering results of dMRI. Furthermore, the color intensity, reflecting the strength of the fMRI information, shows greater coherence along the streamlines within clusters. This enhanced coherence indicates that our approach maintains stronger functional alignment compared to the other three methods, yielding more consistent clustering and functional coherence across the data. These findings underscore the efficacy of our method in achieving both high-quality clustering and more functionally meaningful groupings.

\subsubsection{Visualization of embedding space}
To visually demonstrate the superior performance of our clustering method at the feature level, we employed the Uniform Manifold Approximation and Projection (UMAP) [37] to project 10-dimensional embeddings from the dMRI information view into a 2-dimensional space. Furthermore, the final clustering label will represent how the overall clustering framework divides fibers into different clusters.

In Figure \ref{embedding}, distinct colors represent different clusters, and each point corresponds to a single fiber. The key difference, highlighted by the red circle in the figure, illustrates the overall performance of our clustering framework. We also visualize the corresponding fibers of the red circle embedding areas in the right of the embedding visualization. The blue fibers included in our cluster indicates that our approach achieves better clustering results, due to the additional information and improvements made to the overall framework.  In the figure, with similar embedding, our method achieves better clustering results in some fibers that the baseline failed to assign appropriate labels. This figure demonstrates the enhancement in feature embedding capabilities of our model, showing that it effectively makes data points within the same cluster closer in the lower-dimensional representation and performs more accurate label assignment. This improvement in embedding ability further reinforces the performance of our method in distinguishing and clustering related data points more accurately.

\section{DISCUSSION}
In this work, we introduced DMVFC, a novel deep learning framework that effectively integrates multimodal dMRI and fMRI data to enhance WM parcellation. Unlike traditional methods relying primarily on geometry, DMVFC leverages microstructural (FA) and functional (BOLD) information, leading to improved functional and anatomical coherence of fiber clusters. Our experiments demonstrate DMVFC's superior performance over state-of-the-art methods in achieving functionally homogeneous and geometrically consistent WM parcellation. Ablation studies confirmed the positive impact of our centroid initialization strategy and highlighted the synergistic benefits of integrating both fMRI and FA information, often yielding improvements beyond additive effects. Crucially, DMVFC showed high clustering consistency across subjects, with significantly smaller pairwise distances between representative pathways from different subjects compared to intra-cluster distances, underscoring its effectiveness and robustness in producing reproducible WM parcellations. 
DMVFC's framework also offers substantial flexibility for future extensions. Other diffusion measures like NODDI parameters, task-based fMRI, T1-weighted MRI, and Quantitative Susceptibility Mapping (QSM) could be incorporated to provide more comprehensive and biologically detailed WM characterization. While advantageous for richer insights, such integration entails increased data complexity, potentially more sophisticated network architectures, and higher computational demands. Regarding tractography, our method utilizes TractSeg for bundle delineation. Although DMVFC is designed to be robust to fiber representation variations, different algorithms can yield distinct fiber characteristics. Future work will systematically evaluate DMVFC's performance and robustness across various tractography methods to assess how fiber generation differences propagate through our framework. DMVFC uniquely integrates complementary dMRI and fMRI insights to overcome limitations of geometry-centric approaches. By achieving superior functionally consistent clustering and high inter-subject reproducibility, DMVFC provides a powerful new tool for understanding the intricate structure-function relationship in WM, with significant implications for neuroimaging research and potential clinical applications.

\section{CONCLUSION}
In this paper, we present a novel deep-learning framework for fiber clustering that integrates dMRI, fMRI, and FA information, effectively combining both geometric and functional data from WM fiber tracts. Through extensive experiments, we identified an optimal method for merging the geometric characteristics of fibers with BOLD signals along the WM fibers, enhancing clustering performance. The proposed framework has been rigorously validated, demonstrating its functional and structural consistency. By leveraging multimodal data, our framework underscores the substantial potential of WM BOLD signals in improving fiber clustering accuracy and provides a powerful tool for advanced neuroimaging analysis.

\section*{Acknowledgements}
This work is in part supported by the National Key R\&D Program of China (No. 2023YFE0118600), and the National Natural Science Foundation of China (No. 62371107).
\bibliographystyle{unsrt}
\bibliography{cite}

\end{document}